\pdfoutput=1

\documentclass[article,pre,10pt,twocolumn,longbibliography]{revtex4-1}
\usepackage{microtype}
\usepackage{graphicx}
\usepackage[colorlinks,linkcolor=blue,urlcolor=blue,citecolor=blue]{hyperref}

\usepackage{caption}
\usepackage{subcaption}
\usepackage{mathptmx}
\usepackage{times}
\usepackage{epsfig,amsopn}
\usepackage{graphicx}
\usepackage{bm,amsmath,amssymb}
\usepackage{natbib}
\usepackage{braket}
\usepackage{float}
\usepackage{enumerate}
\usepackage{hyperref}
\usepackage{comment}
\allowdisplaybreaks[1]
\usepackage{tabularx}

\def\be{\begin{equation}}
\def\ee{\end{equation}}
\def\bse{\begin{subequations}}
\def\ese{\end{subequations}}
\def\bcs{\begin{cases}}
\def\ecs{\end{cases}}
\def\bea{\begin{eqnarray}}
\def\eea{\end{eqnarray}}

\newcommand{\eref}[1]{Eq.~\eqref{#1}}%
\newcommand{\fref}[1]{Fig.~\ref{#1}}%
\newcommand{\Aref}[1]{Appendix}%

\newcommand{\opunit}{\textrm{1}\kern-0.22em\textrm{l}}

\def\la{\langle}
\def\ra{\rangle}
\def\l{\left}
\def\r{\right}
\def\s{\sum}
\def\b{\beta}
\def\db{\Delta}

\begin{document}


\title{Inverse transitions and disappearance of the $\lambda$- line in the asymmetric random field Ising and Blume-Capel models}

\author{{\normalsize{}Santanu Das$^{1, 2}$}
{\normalsize{}}}
\email{santanudas@niser.ac.in}

\author{{\normalsize{}Sumedha$^{1, 2}$}
{\normalsize{}}}
\email{sumedha@niser.ac.in}

\affiliation{\noindent $^{1}$School of Physical Sciences, National Institute of Science Education and Research, Jatni 752050, India}

\affiliation{\noindent $^{2}$Homi Bhabha National Institute, Training School Complex, Anushakti Nagar 400094, India}

\begin{abstract}

We report on reentrance in the random field Ising and Blume-Capel models, induced by an asymmetric bimodal random field distribution. The conventional continuous line of transitions between the paramagnetic and ferromagnetic phases, the $\lambda$-line, is wiped away by the asymmetry. The phase diagram, then,  consists of only first order transition lines that always end at ordered critical points. We find that while for symmetric random field distributions there was no reentrance, the asymmetry in the random field results in a range of temperatures for which magnetisation shows reentrance. While this does not give rise to an inverse transition in the Ising model, for the Blume-Capel model, however, there is a line of first order inverse phase transitions that ends at an inverse ordered critical point. We show that the location of the inverse transitions can be inferred from the ground state phase diagram of the model.

\end{abstract}

\date{\today}
\maketitle


\section{Introduction}
Inverse transitions are an unusual class of phase transitions where the ordered phase has more entropy than the disordered phase and hence occurs at a higher temperature \cite{greer:00}. This entropy driven phase reentrance of the ordered phase is widely observed \cite{schupper:05}. Examples include ferroelectric thin films \cite{nahas:20}, perpendicularly magnetized ultrathin ferromagnetic 
flims \cite{portmann:03, saratz:10, saratz2:10}, anisotropic dipolar magnets \cite{brooke:99}, polymer systems such as Poly(4-methyl-1-pentene) \cite{rastogi:93, rastogi:99}, the solutions of cyclodextrin, water and methlpyridine \cite{plazanet:04,angelini:08}, inverse melting between lattice and disordered vortex phase in high-temperature superconductors \cite{avraham:01} and shear thickening in glasses and granular systems \cite{sellitto:05}. 

Models with spin-$1$ variables like the Ghatak-Sherrington model have been found to exhibit inverse transition (IT) in some recent studies \cite{schupper:04, sellitto:06, sellitto:05, paoluzzi:10, crisanti:05, leuzzi:07, morais:13, leuzzi:11, thomas:11, costa:10, ferrari:11, morais:12, coto:19, erichsen:11}. These studies have focussed on models with a glassy phase and random bond interactions, where it is expected that frustration and disorder allows for a possibility of inverse freezing (a glass to liquid transition on cooling). Reentrance is also seen in dipolar long and short range models with asymmetric random interaction and Gaussian random fields \cite{andresen:13}. However, in general it is expected that random fields will suppress the IT \cite{morais:13}.

In this work, we study the random field Ising model (RFIM) and Blume-Capel model (RFBCM) with ferromagnetic interactions and an asymmetric bimodal distribution (BD) for the random field. These models do not have a glass phase. Also, the models with the symmetric BD for the quenched random fields have no ITs \cite{kaufman:90, santos:18}. Any asymmetry in the random field  distribution is expected to make the system less random and hence no ITs are expected. In this paper, we undertake an expansive study of the infinite range RFIM and RFBCM with asymmetric BD and report a number of interesting results. Infinite range interaction models usually belong to the same universality class as the mean-field models with fixed coordination number. Generically, we find that even an infinitesimal asymmetry changes the phase diagram non-trivially. Interestingly, there is a line of inverse first order transitions in the phase diagram of the asymmetric RFBCM.  While there have been some studies of these models with symmetric distributions \cite{kaufman:90, santos:18, mukherjee:22, fytas:08}, asymmetric distributions have hithertho been studied for the  RFIM \cite{maritan:91, hadjiagapiou:10, swift:94}. Disorder distribution is typically asymmetric in real experiments \cite{maritan:91}. We find that asymmetric RFBCM shows first order ITs similar to those seen in experiments that display inverse melting\cite{rastogi:93, rastogi:99, plazanet:04, angelini:08}.

For symmetric BD the RFIM has a line of continuous transitions ($\lambda$-line) that meets a line of first order transitions at a tricritical point (TCP) \cite{aharony:78}. We find that even a slight asymmetry wipes away the $\lambda$-line and the TCP in the RFIM. We instead find a phase diagram consisting of a line of first order transitions that ends at a critical point. The magnetization ($m$) is non zero at this point and hence we call this an ordered critical point (OCP) \cite{bep}. The location of OCP to a good approximation is determined by the location of the first order transition in the ground state phase diagram of the model. Hence even at finite temperature ($T$) the phase diagram is dominated by the random field disorder. 

The fluid separation in porous media is considered a good realization of RFIM \cite{wong:90,bonnet:08,aubry:14}. The results from experiments on these models found the value of the  order-parameter exponent to be closer to the value for the pure Ising Model rather than for the RFIM with symmetric random field distribution \cite{wong:90}. It was suggested that these experiments should be compared with the asymmetric RFIM \cite{maritan:91}. In more recent experiments it is shown that they exhibit out of equilibrium disorder-driven behaviour similar to the athermal non-equilibrium RFIM \cite{kierlik:01,bonnet:08}. Consistent with the experiments, we find that the value of the exponent near an OCP is the same as the pure Ising critical point. 


Another interesting observation is the non-monotonic behaviour of $m$ as a function of $T$ for both asymmetric RFIM and asymmetric RFBCM. We find that for the values of the parameters close to an OCP, $m$ can become non-monotonic. Though in the absence of the crystal field ($\Delta$), there is no IT in these models. We show that for RFBCM for a range of $\Delta$, $m$ jumps to a higher value on increasing $T$. The system has a first order IT which we show is entropy driven. The magnitude of the jump decreases with increasing $T$ and the line of first order IT ends at an inverse OCP. We hence report a mechanism for ITs which crucially depends on the asymmetry of the disorder distribution. This is an inverse melting transition since the system goes from a less ordered state to a more ordered state on increasing $T$.  We also find that the RFBCM has two first order transitions with increasing $T$ for a narrow range of parameters: first from a less ordered state to a more ordered state and then again to a less ordered state, similar to the two first order transitions observed in recent experiments involving solutions of cyclodextrin, water and methlpyridine \cite{angelini:08}. The RFBCM also shows a reentrance in the quadrupole moment ($q$) for some range of the  parameters. We show that the ground state phase diagram crucially determines the phase-diagram at finite $T$. 

\section{Model} 
The Hamiltonian for the infinite range RFIM and RFBCM can be written as
\bea
\mathcal{H}
= - \frac{1}{2 N} \l( \s_{i=1}^N s_i \r)^2 + \Delta \s_{i=1}^N s^2_i 
- \s_{i=1}^N h_i s_i
\label{H(C_N)}
\eea
here $s_i=\pm 1$ and $s_i=0,\pm 1$ for RFIM and RFBCM respectively. The  crystal field is represented by $\Delta$. It is $0$ for RFIM. The RFBCM with $s_i=0,\pm 1$ and $\Delta=0$ has a behaviour which is similar to RFIM with $s_i=\pm1 $. We hence also call it RFIM with $s=1$.

The magnetic field $h_i$ associated with each site is an independent and identically distributed (i.i.d) random variable taken from the BD of the form 
\be
Q (h_i) =
r \delta(h_i - h_0) + (1 - r) \delta(h_i + h_0),  
\label{rf_dist_bim}
\ee 
with bias $r$ and strength $h_0$. The above distribution is asymmetric when $r \neq 1/2$. We take $h_0>0$ and consider $r \in [1/2, 1]$. 

The probability of a spin configuration $C_N$ with magnetisation $x_1=\sum_i s_i/N$ and quadrupole moment $x_2 = \sum_i s_i^2/N$ satisfies large deviation principle (LDP), i.e, $P(C_N : x_1,x_2)  \sim e^{-N I(x_1,x_2)}$. $I$ is a rate function that can be calculated using large deviations. The free energy of the system is then the infimum of $I$ with respect to $x_1$ and $x_2$. It is hence enough to consider only the fixed points of $I$ to write the generalized free energy functional of the model. We hence obtain an expression for the free energy functional of the model with quenched random fields (see \cite{mukherjee:22} for details) as
\be
\widetilde{f}(x_1) = \frac 12 \b x_1^2 - \l\la \log \l( c + 2 e^{- \b \db} \cosh \b(x_1 +h_i) \r) \r \ra_{\{h_i \}},
\ee
where $\b = 1/T$, $c=0$ for RFIM and $c=1$ for RFBCM. The value of $x_1$ that minimises $\widetilde f(x_1)$ is the magnetisation $m$ and the quadrupole moment $q = \frac{1}{\beta} \partial \widetilde{f}(x_1)/\partial \Delta|_{x_1=m}$. These are given by

\be
m = \l\la  \frac{ 2 e^{- \b \db} \sinh \b(m +h_i)}{ c + 2 e^{- \b \db} \cosh \b (m + h_i)} \r \ra_{\{h_i \}},
\label{mag_app}
\ee
and the quadrupole moment 
\be
q = \l\la \frac{ 2 e^{- \b \db} \cosh \b(m +h_i)}{ c + 2 e^{- \b \db} \cosh \b (m + h_i)} \r \ra_{\{h_i \}}.
\ee
$\langle \rangle_{\{h_i\}}$ represents the average over random field distribution.


\begin{figure}[t]
\centering{\includegraphics[width=1.0\hsize]{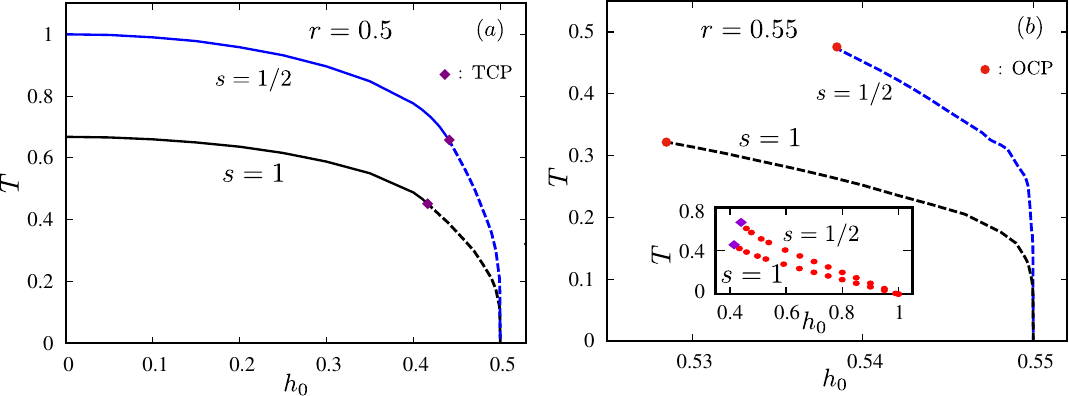}
\caption{Phase diagram in $(T-h_0)$ plane for RFIM $(s = \pm 1)$(blue) and its spin-$1$ variant $(s = 0, \pm 1)$ (black) for  (a) symmetric BD ($r=0.5$) and (b) asymmetric BD ($r=0.55$). Solid lines are the lines of continuous transitions and the dashed lines are the lines of first order transitions. Rhombus(purple) represents the TCP and circle(red) represents an OCP. Inset of (b) plots the locus of the TCP (rhombus) and the OCP (circle) in the $(T-h_0)$ plane for $1/2 \leq r \leq 1 $. With increasng $r$, the OCP occurs at a lower value of $T$ and higher value of $h_0$.}\label{pd_1}}
\end{figure}



\section{ RFIM for $s=1/2$ and $s=1$}  
The phase diagram of the RFIM for symmetric BD is known since long \cite{aharony:78, schneider:77}. It has a line of continuous transitions between ordered and disordered phases for the weak disorder strength ( low $h_0$) that ends at a TCP. On further increasing $h_0$, there is a line of first order transitions that ends at $h_0=1/2$ and $T=0$. The qualitative phase behaviour remains unchanged for spin-$1$ system in the absence of $\Delta$ (see \fref{pd_1}(a)). 

\begin{figure}[t]
\centering{\includegraphics[width=1.0\hsize]{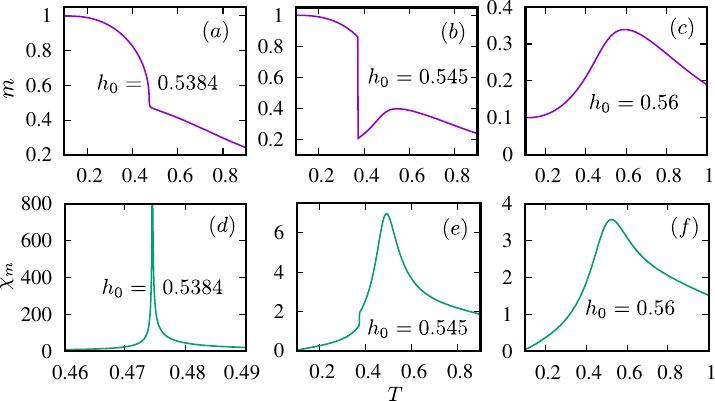}
\caption{Magnetisation($m$) and magnetic susceptibility($\chi_m$) for RFIM with $s=1/2$ and $r=0.55$ is plotted at the OCP ((a)and (d)), at a point along the first transition line ((b) and (e)) and  for $h_0$ near the first order transition line with reentrance in $m$ ((c) and (f)).}\label{ms}}
\end{figure}

Interestingly, we find that for an asymmetric BD (\eref{rf_dist_bim}), asymmetry in the distribution wipes out the line of continuous transitions along with the TCP. The phase diagram only has a line of first order transitions that starts at $h_0=r$ and $T=0$ and ends at an OCP. As $r$ deviates from $1/2$ and approaches $1$, the OCP occurs at a lower value of T and  approaches $0$ as $r \rightarrow 1$ (see \fref{pd_1}(b)). Since $m$ is finite at an OCP, to find the co-ordinates of the OCP we equate the first three derivatives of $\widetilde{f}(x_1)$ to $0$. The meeting point of the solution of the three equations, for a given $r$, $\db$ and $h_0$, gives the co-ordinates of the OCP \cite{domb:00}. In \fref{ms} for $s=1/2$ we have plotted magnetisation and susceptibility at three different points in the phase diagram : at OCP, at a point on the line of first order transitions between $m \approx 1$ and $m \approx 2 r-1$ and for a point near the first order line where there is no transition but magnetisation $m$ is non-monotonic.  Similar behaviour occurs for $s=1$ as well. We find that both for $s=1/2$ and $s=1$, $m$ shows a non-monotonic dependence on $T$ for any $r > 1/2$ and $h_0 > r$. The degree of non-monotonicity is maximum when $r$ is close to $1/2$ and $h_0$ is just above $r$. 

\begin{figure}[t]
\centering{\includegraphics[width=1.0\hsize]{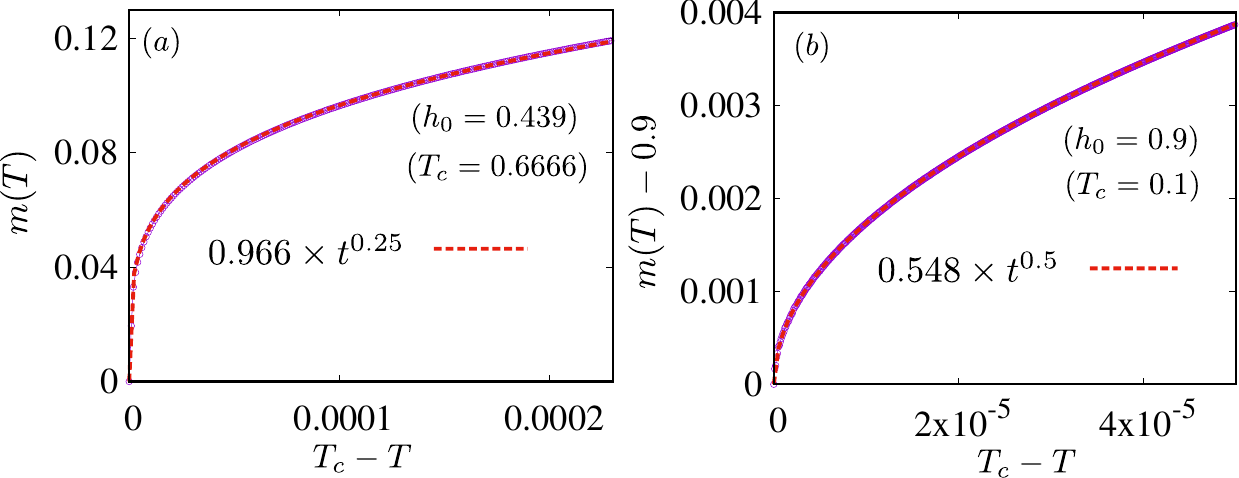}
\caption{Magnetization $(m \sim t^{\beta})$ versus the scaled temperature $t = T_c- T$ for RFIM is plotted in the vicinity of TCP and OCP  in (a) and (b) for $r = 0.5$ and $0.9$ respectively. The points are the numerical value of magnetisation $m$ and  the red dashed line is the scaling fit in both the cases.}
\label{scaling}}
\end{figure}

We find that the OCP lies in the critical Ising universality class and $m$ scales with exponent $\beta =1/2$ near an OCP as $T$ increases. On the other hand, $\beta = 1/4$ for a TCP. This is verfied in Fig \ref{scaling}, where we contrast the scaling of magnetisation near a TCP and an OCP by taking symmetric and asymmetric BD respectively.

\section {RFBCM and the reentrance transition}
\begin{figure}[t]
\centering{\includegraphics[width=1.0\hsize]{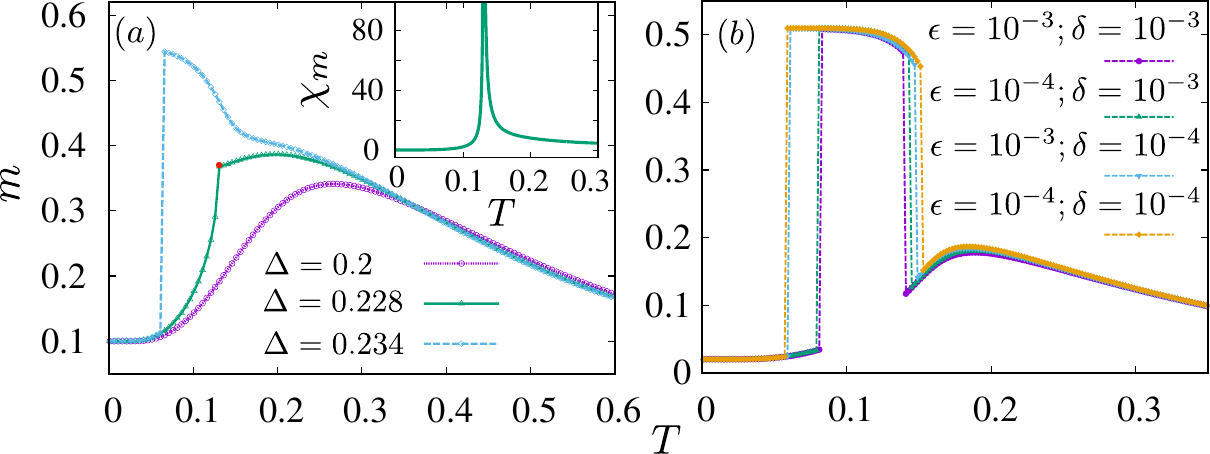}
\caption{(a) $m$ versus $T$ for RFBCM for $r = 0.55$ and $h_0 = 0.56$ for different $\Delta$. The red dot is the OCP. In the inset the susceptibility for the critical $\Delta = 0.228$ (OCP) is plotted. (b) $m$ versus $T$ for RFBCM for $r = 0.51$ with small $\epsilon$ and $\delta$ where $h_0 = r + \epsilon$ and $\Delta = h_0 - (3r-1)/2 -\delta$.}
\label{re-ent_1}}
\end{figure}

For spin-$1$, on the introduction of the $\Delta$, i.e. for RFBCM we find that there is a first order reentrance transition for the  asymmetric BD for a range of $\Delta$. The transition becomes a continuous reentrance transition at $(\Delta_c,T_c)$ that depends on the values of $r$ and $h_0$ (see Fig. \ref{re-ent_1}(a)). We also find that depending on the value of $r$, there is also a possibility of a second first order transition from a more ordered to a less ordered state in the model (see Fig. \ref{re-ent_1}(b)). 

\begin{figure}[t]
\centering{\includegraphics[width=0.95\hsize]{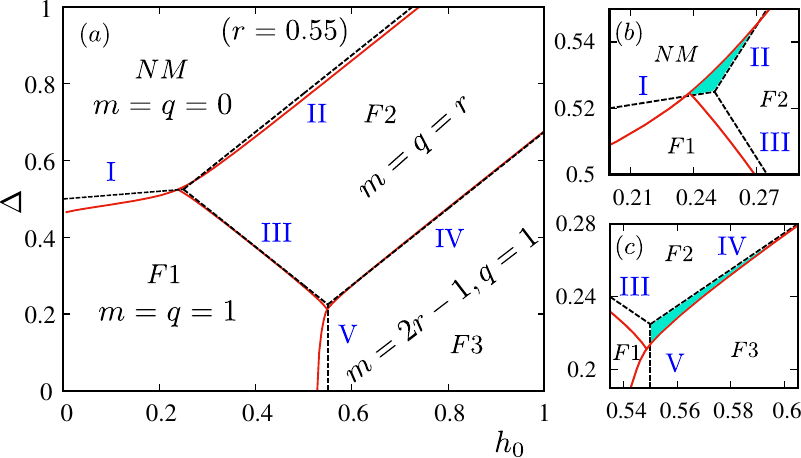}
\caption{(a) The $T=0$ phase diagram in $(\Delta-h_0)$ plane for $r = 0.55$ with three ferromagnetic phases: $F1$, $F2$, $F3$ and a non-magnetic phase $NM$. Black dashed lines are the lines of first order transitions between the two neighboring phases at $T=0$ given by {\MakeUppercase{\romannumeral 1} : $\Delta = 1/2 - (1-2r) h_0$}, {\MakeUppercase{\romannumeral 2} : $\Delta = h_0 + r/2$}, {\MakeUppercase{\romannumeral 3} : $\Delta = (1+r)/2 - h_0$}, {\MakeUppercase{\romannumeral 4} : $\Delta = h_0 -(3r-1)/2$} and {\MakeUppercase{\romannumeral 5} : $h_0 = r$}. Solid red lines are the projection of the OCPs in $(\Delta-h_0)$ plane. In (b) and (c) we enlarge the vicinity of the two triple points. The shaded part shows the range of parameters for which the IT in $m$ occurs.}\label{pd_2}}
\end{figure}

\begin{figure}[t]
\centering{
\includegraphics[width=1.0\hsize]{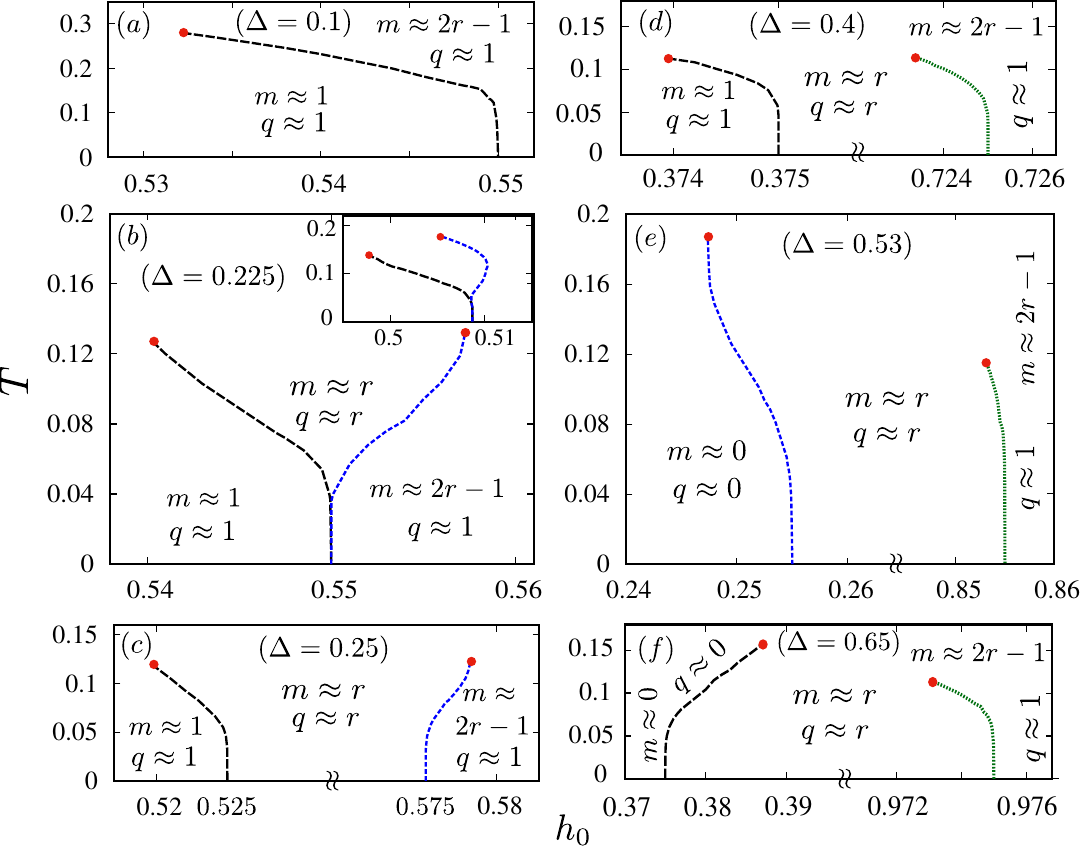}
\caption{Different phase diargams for RFBCM for different ranges of $\Delta$ in $(T-h_0)$ plane for $r = 0.55$. (a) $-\infty < \Delta \le 0.211$, (b) $0.211 < \Delta \le 0.225$, (c) $0.225 < \Delta \le 0.296$, (d) $0.296 < \Delta \le 0.524$, (e) $0.524< \Delta < 0.545$ and (f) $\Delta > 0.545$. Black lines are the lines of usual first order transition, Blue lines are the lines of first order IT in $m$ and green lines are the lines of first order IT in $q$. Red dots are the OCPs.  Inset in (b) shows the phase diagram for $r=0.51$, where the first order transition line (blue) bends back, giving rise to two first order transitions as a function of $T$ for fixed $h_0$. }\label{pd_3}}.
\end{figure}

To understand the phase behaviour at finite temperature, we first study the ground state $(T = 0)$. In the ground state, the disorder averaged energy is given by $\underset{m}{min}\; \phi(m)$, where $\phi(m) = \underset{\b \to \infty}{lim} \b^{-1} \widetilde{f}(m)$. We find that the ground state ($T=0$) phase diagram of the RFBCM has four phases (three ferromagnetic phases $F1$,$F2$, and $F3$ and one nonmagnetic phase $NM$). These phases are separated by the lines of first order transitions (see \fref{pd_2}). These transitions can be understood by looking at the configurational entropy of these states. For example, the phases $F2$ and $F3$ have same configurational entropy as in both phases spins take two values: in $F3$ $\pm 1$ and in $F2$ $0,1$. As $\Delta$ increases, $0$ spins become more favourable energetically and first there is a transition from $F3$ to $F2$ and finally to $NM$ (phase with all spins $0$). As $T$ increases, each point on these first order transition lines changes its position and ends at an OCP. The phase diagram of the model in the $(T-h_0)$ plane, for different ranges of $\Delta$ for $r=0.55$ is shown in \fref{pd_3}. We find that the finite $T$ phase diagrams only have lines of first order transitions and OCPs. This is very different from the phase diagrams for RFBCM with symmetric bimodal and trimodal distributions \cite{kaufman:90, santos:18, mukherjee:22}. For symmetric distributions, the phase diagrams consist of lines of first and second order transitions and various multicritical points.

Depending on the strength of the crystal field, there are six different finite temperature phase diagrams for the asymmetric BD. The phase diagram for the asymmetric BD for $r=0.55$ in $(T-h_0)$ plane for $\Delta < \Delta_1(= 0.211)$ is similar to the $\Delta = 0$ case : single first order line of transitions separates $m \approx 1$ from $m \approx 2 r-1$ and ends at an OCP (see \fref{pd_3}(a)). For $\Delta > \Delta_1$, interestingly we find two lines of first order transitions, both ending at OCPs. For $\Delta_1 < \Delta < \Delta_2(=0.296)$, one of them corresponds to the usual first order transition from a more ordered to a less ordered state (shown in black) and the other is a line of first order IT (shown in blue) between states with $m \approx 2 r -1$ and $m \approx r$ (see \fref{pd_3}(b) and (c)). On further increasing $\Delta$, the reentrance transition in $m$ changes to a reentrance transition in $q$ as shown by the green lines in \fref{pd_3}(d), (e) and (f). For $0.525 < \Delta < 0.545$, near the second triple point in the ground state (\fref{pd_2} (b)), IT occurs for both $m$ and $q$ as shown in \fref{pd_3} (e).

We projected the OCPs onto the ground state phase diagram of the model and identified the region in the $(\Delta-h_0)$ plane where the IT occurs. Corresponding to the first order line of transitions in the ground state phase diagram, we find a line of projection of OCPs in the $(\Delta-h_0)$ plane (\fref{pd_2}(a)). When this line of projections of OCPs enters into either $F3$ or the $NM$ phase, there is a region in the $(\Delta-h_0)$ plane where the reentrance transition takes place. For $r=1/2$ this region shrinks to zero and there is no reentrance. In \fref{pd_2}(b) and (c) the range of $(\Delta, h_0)$ for which there is a IT in $m$ is shown shaded for $r = 0.55$.  The reentrance region at first increases with $r$ and then shrinks as $r \rightarrow 1$. 

To find the region in the phase diagram where reentrance occurs, we fixed $h_0 \gtrsim r$ and gradually increased $\Delta$. For example, for $r = 0.55$ and $h_0 = 0.56$, we find first order reentrance transition for $ 0.228 \leq \Delta \leq 0.235$ ( \fref{re-ent_1}(a)). As $\Delta \rightarrow 0.228$ there is still a reentrance, but without a jump in $m$. We find that this point is in fact an OCP. Inset of \fref{re-ent_1}(a) shows the divergence of the  magnetic susceptibility at the OCP.

We also find that if we take $\Delta$ and $h_0$ very close to the triple-point of the $T=0$ phase diagram for $r \gtrsim 1/2$, then there are two first order transitions with the increase of $T$ (see inset of \fref{pd_3} (b)). For example, for $r = 0.51$, when we set $h_0 = r + \epsilon$ and $\Delta = h_0 - (3r-1)/2 - \delta$ (where $\epsilon$ and $\delta$ are small) in the vicinity of the triple-point, we observe two first order transitions. For $r = 0.51$ this is shown in \fref{re-ent_1}(b). This double first order transition is similar to the one seen in experiments with solutions of cyclodextrin, water and $4$-methylpyridine which go from low-density-liquid to high-density-liquid to low-density-liquid on increasing $T$ via two first order transitions \cite{angelini:08}.   

If instead of fixing $h_0$, we fix $\Delta \gtrsim (1+2r)/4$ then also we find a region in the phase diagram where the reentrance transition occurs. In fact the phase diagrams in $(T-\Delta)$ plane are similar to phase diagrams in the $(T-h_0)$ plane.

\begin{figure}[t]
\centering{
\includegraphics[width=1.0\hsize]{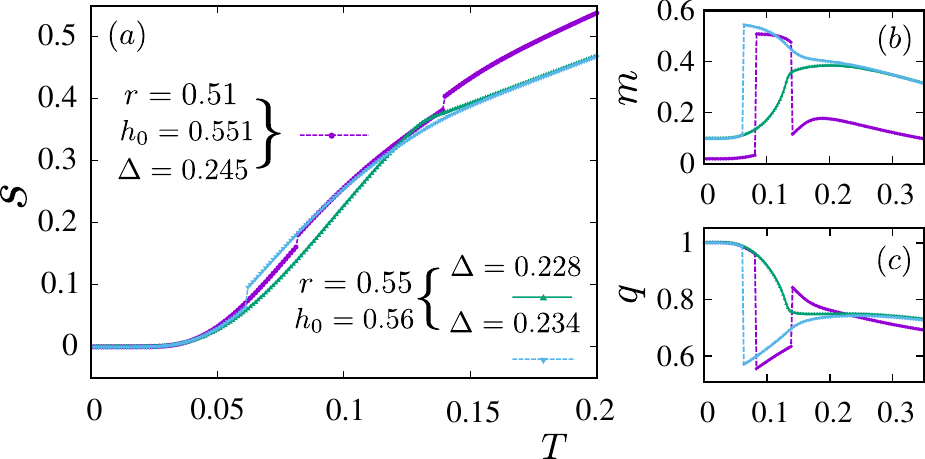}~
\caption{Plots of entropy ($s$), $m$ and $q$ as a function of $T$ for OCP (solid green line), near IT when there is only one first order transition (dotted blue) and for the case where there are two first order transitions (dotted purple).}\label{ent}
}
\end{figure}

\section{Concluding Remarks}
We showed that the asymmetry in the random field distribution results in a non-monotonic behaviour of the order parameter in the ferromagnetic models with quenched random fields that becomes an IT on the introduction of $\Delta$. To understand this let us look at the $T=0$ phase diagram again. At $T=0$, there is a residual $m$ of order $2r-1$ at low $\Delta$ and high $h_0$ for asymmetric BD (phase $F3$ in Fig. \ref{pd_2}). For symmetric BD the $F3$ becomes a paramagnetic phase and if the $\Delta$ and $h_0$ are chosen such that the system is in this state at $T=0$, then the system continues to stay in that state with $m=0$ on increasing $T$ as $m=0$ maximises entropy. On the other hand for $r \gtrsim 1/2$ and $(\Delta,h_0)$ very close to the triple point in \fref{pd_2}(c), $m$ increases as $T$ increases and then jumps to $r$ at the IT transition point. The entropy ($s$) also jumps at that point (Fig. \ref{ent}). Since an infintesimal amount of asymmetry can give rise to IT, it is possible that the topology of finitely connected graphs with heterogeneous degree distribution can induce that asymmetry and give rise to topology induced IT as seen in some studies \cite{erichsen:17, martino:12}. 

The value of $\Delta$ and $h_0$ at which the IT occurs is close to the triple points in the ground state phase diagram.
The infinite range pure Blume Capel model ($h_0=0$) gives the true behaviour of the model in finite dimensions. Also numerical study of the Ghatak-Sherrington model in three dimensions have reported first order inverse freezing transition  \cite{paoluzzi:10, leuzzi:11}. We expect our result of the appearence of IT near the triple point of the ground state will hold in finite dimensions for RFBCM as well. 

The absence of IT for symmetric distribution has also been reported for continuous spin models with random fields like the random field $XY$ model \cite{sumedha:22, lupo:22}. We expect that the asymmetry in the distribution should induce reentrance in the case of random field models with continuous spin as well. 

For RFIM it was conjectured that if the phase diagram has a TCP for the symmetric distribution, it will change to critical end point for any infintesimal asymmetry \cite{maritan:91,swift:94}. Presence of the critical end point implies that the $\lambda$-line is still present in the phase diagram.
In contrast, for the asymmetric BD defined via \eref{rf_dist_bim}, we find that the $\lambda$-line and the TCP both disappear completely and there is an OCP instead of a TCP in the phase diagrams. 

We also studied the asymmetric Gaussian random field distribution. The $\widetilde{f}$ of the asymmetric Gaussian RFIM is the same as that of the symmetric Gaussian RFIM in an external field of strength equal to the bias in the distribution. Since symmetric Gaussian RFIM in an external field has finite $m$ at all $T$ that gradually goes to 0 without a phase transition, asymmetric Gaussian RFIM also has no phase transition. Another interesting distribution is the double peaked asymmetric Gaussian distribution. This we expect will have the  similar phase diagrams as the asymmetric BD as long as the variance of the distribution is not large.
 






\end{document}